\documentclass[twocolumn,noshowpacs,amsmath,amssymb]{revtex4-2}
\usepackage{graphicx}
\usepackage{amsfonts}
\usepackage{dcolumn}
\usepackage{bm}
\usepackage{color}
\usepackage{CJK}
\usepackage{hyperref}
\usepackage{booktabs}

\begin{document}
	
\newcommand{\gin}[1]{{\bf\color{blue}#1}}
\def\bc{\begin{center}}
\def\ec{\end{center}}
\def\bea{\begin{eqnarray}}
\def\eea{\end{eqnarray}}
\newcommand{\avg}[1]{\langle{#1}\rangle}
\newcommand{\Avg}[1]{\left\langle{#1}\right\rangle}

\title{\textcolor{blue}{Non-equilibrium} random walks on multiplex networks}
\author{Feng Huang$^{1}$}
\author{Hanshuang Chen$^{2}$}\email{chenhshf@ahu.edu.cn}

\affiliation{$^1$School of Mathematics and Physics \& Key Laboratory of Architectural Acoustic Environment of Anhui Higher Education Institutes \& Key Laboratory of Advanced Electronic Materials and Devices, Anhui Jianzhu University, Hefei 230601, China \\ $^{2}$School of Physics and Optoelectronic and Engineering, Anhui
	University, Hefei 230601, China}

\date{\today}
	
\begin{abstract}	
We \textcolor{blue}{introduce a non-equilibrium} discrete-time random walk model on multiplex networks, in which at each time step the walker first undergoes a random jump between neighboring nodes in the same layer, and then tries to hop from one node to one of its replicas in another layer. We derive the so-called supra-Markov matrix that governs the evolution of the occupation probability of the walker. The occupation probability at stationarity is different from the weighted average over the counterparts on each layer, unless the transition probabilities between layers vanish. However, they are approximately equal when the transition probabilities between layers are very small, which is given by the first-order degenerate perturbation theory. Moreover, we compute the mean first passage time (MFPT) and the graph MFPT (GrMFPT) that is the average of the MFPT over all pairs of distinct nodes. Interestingly, we find that the GrMFPT can be smaller than that of any layer taken in isolation. The result embodies the advantage of global search on multiplex networks.  
\end{abstract}
	\maketitle
	
\section{Introduction}

In the past two decades, we have witnessed the power of network science on modeling dynamical processes in complex systems made of large numbers of interacting elements \cite{newman2018networks,RMP08001275,PRP06000175}.
However, the recent availability of massive data sets of social, technological and biological systems has suggested that many real-world complex systems are usually composed of interwined multilayer networks \cite{boccaletti2014structure}. As an example, transportation networks between cities are formed by different types of networks, such as highway network, railway network, and airline network, etc. Another example is represented by online social networks where each layer corresponds to a different social structure (e.g., Facebook and Twitter) and users play the role of nodes. It has been recognized that the multilayer networks can not only induce some novel features different from the single-layer networks, such as complexity, diversity and fragility \cite{Nature2010,PhysRevLett.105.048701,PhysRevLett.107.195701,PhysRevLett.109.248701}, but also offer a better understanding for dynamical behaviors, including epidemic spreading \cite{NatPhy12.901.2016,Arruda2018physrep}, vaccination \cite{wang2016statistical}, synchronization \cite{PhysRevLett.112.248701,PhysRevLett.114.038701}, evolution of cooperation \cite{wang2015EPJB}, opinion formation \cite{PhysRevE.89.062818}, transportation \cite{PhysRevLett.116.108701,PhysRevLett.120.068301}. Multiplex is a particular type of multilayer network in which each agent participates in different layers simultaneously, just as our previous example in the case of online social networks. A mathematical framework has been developed to generalize several important network descriptors and dynamical processes \textcolor{blue}{on multilayer networks}. \textcolor{blue}{As a special case, it can be used for multiplex analysis \cite{PhysRevX.3.041022}}.

In the realm of dynamical processes on networks, random walk is a very simple but important model \cite{PhysRevLett.92.118701,masuda2017random,PRP2014RW}. It not only lies in the heart of many transport processes on networked systems such as the spreading of information or epidemics \cite{RevModPhys.87.925}, human mobility \cite{PhysRevE.86.066116,riascos2017emergence,barbosa2018human}, but also finds a broad range of applications in community detection \cite{rosvall2008maps,zhou2004network,pons2005computing}, ranking and searching on the web \cite{newman2005measure,lu2016vital,kleinberg2006complex,RevModPhys.87.1261}. In this context, two important physical quantities are relevant. One is the occupation probability at stationarity, which quantifies the frequency to visit each
node in the long time limit. The other one is the mean first-passage time (MFPT), which is the average time from one node to another one for the first time \cite{redner2001guide}. First passage underlies a wide variety of important problems like epidemic extinction \cite{WKBReview,PhysRevLett.117.028302}, neuronal firing \cite{tuckwell1988introduction}, consensus formation \cite{PhysRevLett.94.178701}. For a single-layer network, the computations of two quantities are well-established. The stationary occupation probability of the walker at a node is proportional to its degree or strength \cite{PhysRevLett.92.118701,PhysRevE.87.012112}. The MFPT can be calculated by some different approaches, such as the approach through the adjoint equation \cite{van1992stochastic}, the renewal approach combined with spectral decomposition of the transition matrix \cite{PhysRevLett.92.118701}.

\textcolor{blue}{Some existing works have devoted to the super-diffusive behavior of diffusion processes on multiplex networks \cite{PhysRevLett.110.028701,NJP21.035006.2011,PhysRevX.8.031071}, which is related to a structural transition of the multiplex from a decoupled regime to a systemic regime \cite{NatPhys8.717.2018,PhysRevE.88.032807}. However, the diffusive processes are described by a set of linear first-order differential equations, which does not take into the discreteness and stochasticity in the nodal level into account. Many real diffusive processes in networks, such as communication on online social
networks, people commuting on transportation networks, are better modelled by random walks.} \textcolor{blue}{De Domenico} et al. considered different types of random walk on multilayer networks and investigated the time-dependent average fraction of distinct nodes that are visited by walker at least once (in any layer) \cite{PNAS111.8351.2014}. They then examined the coverage as a function of time when some nodes are deleted to model the resilience of multilayer networks to random node failures.  
\textcolor{blue}{Battiston et al. introduced a class of biased random walks on multiplex networks and computed the stationary occupation probability and entropy rate, in which the bias depends on the information of each node at different layers \cite{NJP18.043035.2016}.
Essentially, the work has aggregated all layers to a monoplex network and defined the random walk on the top of it, which can not completely model the random walk on multilayer systems. }
Guo et al. studied L\'evy random walks on multiplex networks and found that the efficiency of such a navigation strategy varies nonmonotonically with an index parameter in L\'evy flights when the interlayer coupling is weak enough \cite{SciRep6.37641.2016}. This result is in contrast to the case in a single-layer network \cite{PhysRevE.86.056110}. \textcolor{blue}{De Domenico et al. proposed a modified dynamics of random walks on multilayer networks, where movements intralayer are Markovian and movements across layers are non-Markovian. Based on the dynamic flows of random walks, they have identified  the community structure of multilayer networks and revealed highly overlapping organization in interconnected systems \cite{PhysRevX.5.011027}.}  \textcolor{blue}{Kuncheva and Montana proposed an algorithm for detecting community structure on multiplex networks. The algorithm is based on a random walk with the transition probabilities depending on the local topological similarity between layers \cite{Kuncheva2015}.}
\textcolor{blue}{Ghavasieh and De Domenico introduced a framework for
functional reducibility to enhance transport phenomena in multilayer systems by coupling layers together with respect to dynamics rather than structure. The framework provide a promising way to reduce diffusion time and optimize noncompact search processes in empirical multilayer systems, without the cost of altering the underlying structure \cite{PhysRevResearch.2.013155}.} \textcolor{blue}{Nasiri et al. proposed a method for link prediction on multiplex networks by using random walk models to collect interlayer similarities \cite{CSF2021.151.111230}. Valdeolivas et al. extended the random walk with restart algorithm to multiplex and heterogeneous networks and explored different layers of physical and functional interactions between genes and proteins \cite{Bioinformatics2019}.} \textcolor{blue}{In a recent work, Bertagnolli and De Domenico introduced a consistent notation and terminology for random walks on multilayer networks. They also extend the framework of diffusion geometry to the realm of multilayer systems, highlighting the dependence of the diffusion space on the interplay between structure and dynamics within and across layers \cite{PhysRevE.103.042301}.} \textcolor{blue}{Gueuning et al. proposed a model of continuous-time random walks on temporal and multiplex networks, where all edges incident to the arrival node of walker are assigned random activation times and the first edge to reach activation is then followed by the random walker \cite{Gueuning2020}. They found that the competition between layers due to the temporality can lead to counter-intuitive phenomena, such as the emergence of a cyclic, rock–paper–scissors precedence. They also explored numerically the impact of the activation mechanism on the coverage of a walker.}

In the present work, we propose a \textcolor{blue}{non-equilibrium} discrete-time random walk model on multiplex networks. At each time step, the walker first performs a random walk between neighboring nodes in the same layer and then tries to switch between replica nodes among different layers. The motivation of the model is to imitate, for example, the information dissemination on online social networks. A user can send a piece of message to one of his friends by Skype, and then his friend further forwards this message by other online social media softwares, such as WeChat, WhatsApp, etc. For the model, we derive the so-called supra-Markov matrix governing the evolution of the occupation probability of the walker at each node. We show that the stationary occupation probability on a multiplex network cannot simply reduced to a conglomerate of additive processes on each single-layer network. Unless the interlayer coupling is weak enough, they are approximately equal to each other from the degenerate perturbation theory. Moreover, we compute the graph mean first passage time (GrMFPT), i.e., the average of MFPT over all pairs of distinct nodes. Interestingly, we find that the GrMFPT can vary monotonically or nonmonotonically with the transition probability between different layers. The GrMFPT can be smaller than that of any layer in isolation, and even that of an aggregate network by all layers. 

\begin{figure}
	\centerline{\includegraphics*[width=0.9\columnwidth]{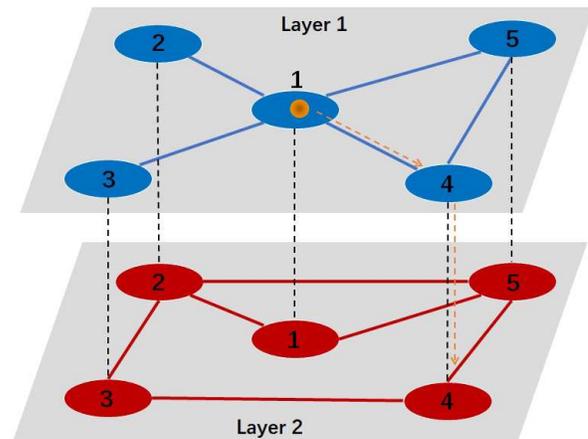}}
	\caption{A multiplex network with $\mathcal{L}=2$ layers. For each layer, the network is described by adjacency matrix $\bm{A}^{(\alpha)}$ with $\alpha = 1, \cdots, \mathcal{L}$. At each time step, a walker jumps randomly from a node to one of its neighboring nodes in the same layer, and then tries to switch between layers in terms of the Markov matrix $\bm{\pi}$.  \label{fig1}}
\end{figure}

\section{Model}\label{sec2}
We consider a walker that performs discrete-time random walks on a multiplex network. The network is consisted of $\mathcal{L}$ layers. Each layer contains the same number of nodes, $N$, and there exists a one-to-one correspondence between nodes in different layers. The topology in each layer is described by an adjacency matrix $\bm{A}^{(\alpha)}$ ($\alpha = 1, \cdots, \mathcal{L}$), whose entries $A_{ij}^{(\alpha)}$ are defined as $A_{ij}^{(\alpha)}=1$ if there is an edge from node $j$ to node $i$ in the $\alpha$th layer, and $A_{ij}^{(\alpha)}=0$ otherwise. For the sake of
simplicity, we consider the case where all connections are undirected and each intralayer network is strongly connected. At each time step, 
the walker first jumps from node $i$ in the $\alpha$th layer (denoted by $i_{\alpha}$) to one of its neighborhood in the same layer, saying $j_{\alpha}$, and then tries to make an interlayer switch (see Fig.\ref{fig1} for an illustration). The transition probability in the same layer can be written as $W_{ij}^{(\alpha)}=A_{ij}^{(\alpha)}/d_{i}^{(\alpha)}$, where $d_{i}^{(\alpha)}$ is the degree of node $i$ at the $\alpha$th layer. The transitions across layers can be described by a Markov matrix $\bm{\pi}$ whose entry $\pi_{\alpha \beta}$ gives the transition probability from the $\alpha$th layer to the $\beta$th layer, where we have assumed that the transition probability between layers is independent of the node's label.

It is obvious that our model is equivalent to standard random walks on a single-layer network when the topologies of all the layers are the same. Our goal is whether our model produces some unexpected results when the topology of each layer is different.
It is also worth to mention that in our model the way to random walks is similar to some previous works \cite{PNAS111.8351.2014,SciRep6.37641.2016,NJP18.043035.2016,JComputSci28.343.2018,SciRep5.9869.2015}, but the definition of the transition matrix is essentially different from the others. In the forthcoming section, we will see that our model in the steady state does not satisfy the so-called detailed balance condition.

\section{Stationary occupation probability}
\subsection{General theory}
Let us denote by $P_{j_{\beta}}(t|i_{\alpha})$ the probability to find the walker at node $j_{\beta}$ at time $t$ providing that the walker has started from node $i_{\alpha}$ at $t=0$. The  master equation governing the time evolution of $P_{j_{\beta}}(t|i_{\alpha})$ reads, 
\begin{eqnarray}\label{eq1}
{P_{{j_{ \beta  }}}}\left( { t |{i_{ \alpha  }}} \right) = \sum\limits_{\gamma  = 1}^\mathcal{L} {{\pi _{\gamma \beta }}\sum\limits_{k = 1}^N {{P_{{k_{ \gamma  }}}}\left( { {t - 1} |{i_{ \alpha  }}} \right)W_{kj}^{\left( \gamma  \right)}} }.  
\end{eqnarray}
Let us indicate with a row vector ${{\bm P}^{\left( \beta  \right)}} = \left( {{P_{{1_{ \beta  }}}}, \ldots ,{P_{{N_{ \beta  }}}}} \right)$ with respect to layer $\beta$, and thus Eq.(\ref{eq1}) can be written as the matrix form, 
\begin{eqnarray}\label{eq2}
{{\bm P}^{( \beta  )}}\left( { t |{i_{ \alpha  }}} \right) = \sum\limits_{\gamma  = 1}^\mathcal{L} {{\pi _{\gamma \beta }}{{\bm P}^{\left( \gamma  \right)}}\left( { t-1 |{i_{ \alpha  }}} \right){\bm{W}^{( \gamma  )}}} .
\end{eqnarray}
Furthermore, we introduce a supra-vector $\bm P = \left( {{{\bm P}^{\left( 1 \right)}}, \ldots ,{{\bm P}^{\left( \mathcal{L} \right)}}} \right) = \left( {{P_{{1_{ 1 }}}}, \ldots ,{P_{{N_{ 1 }}}}, \ldots ,{P_{{1_{ \mathcal{L} }}}}, \ldots ,{P_{{N_{ \mathcal{L} }}}}} \right)$, such that Eq.(\ref{eq2}) can be rewritten as 
\begin{eqnarray}\label{eq3}
\bm P\left( { t |{i_{ \alpha  }}} \right) = \bm P\left( { {t - 1} |{i_{ \alpha  }}} \right) \bm { {W}} ,
\end{eqnarray}
where 
\begin{eqnarray}\label{eq4}
\bm { W} = \left( {\begin{array}{*{20}{c}}
	{{\pi _{11}}\bm{W^{\left( 1 \right)}}}& \cdots &{{\pi _{1 \mathcal{L}}}\bm{W^{\left( 1 \right)}}}\\
	\vdots & \ddots & \vdots \\
	{{\pi _{\mathcal{L}1}}\bm{W^{\left( \mathcal{L} \right)}}}& \cdots &{{\pi _{\mathcal{L}\mathcal{L}}}\bm{W^{\left( \mathcal{L} \right)}}}
	\end{array}} \right).
\end{eqnarray}	
Note that $\bm { W} $ is also a Markov matrix satisfying the sum of entries of each row is always equal to one. We call $\bm { W} $ the supra-Markov matrix, which is borrowed from the supra-Laplacian matrix proposed in \cite{PhysRevLett.110.028701} to model the diffusion process on a multiplex
network, and the normalized supra-Laplacian matrix proposed subsequently in \cite{PNAS111.8351.2014} to model continuous time random walks on a multiplex
network. In the limit of $t\to \infty$, Eq.(\ref{eq3}) gives the stationary equation, 
\begin{eqnarray}\label{eq5}
\bm P\left( \infty  \right) = \bm P\left( \infty  \right)\bm{ W},
\end{eqnarray}	
where we have dropped the conditional probability since in the long time limit 
$\bm P( \infty )$ does not depend on the initial condition. Eq.(\ref{eq5}) implies that $\bm P( \infty )$ is the left eigenvector of $\bm { W} $ corresponding to the unit eigenvalue. Based on $\bm P( \infty )$, we can compute the stationary occupation probability of the walker at node $i$ regardless of the layer, 
\begin{eqnarray}\label{eq5.1}
P_i ( \infty  ) = \sum_{\alpha=1}^{\mathcal{L}} P_{i_\alpha}(\infty).
\end{eqnarray}	

\begin{figure}
	\centerline{\includegraphics*[width=1.0\columnwidth]{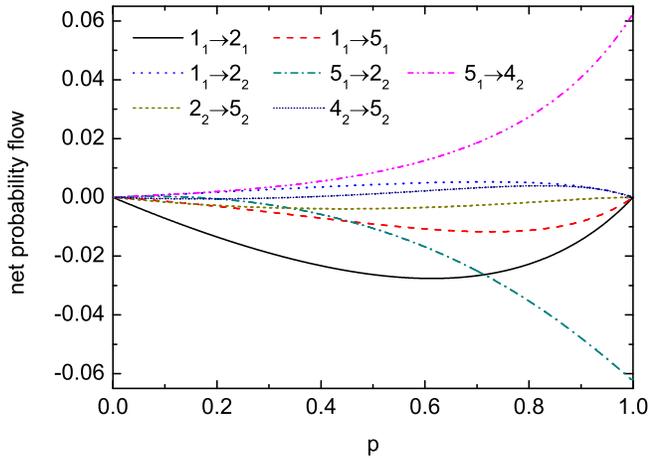}}
	\caption{\textcolor{blue}{The net probability currents at stationarity along seven different reversible transitions on a multiplex network shown in Fig.{\ref{fig1}}. Here the transition probabilities between two layers are set to the same, $\pi_{12}=\pi_{21}=p$. } \label{probflow}}
\end{figure}

\textcolor{blue}{
We should emphasize that our model in the steady state does not satisfy the detailed balance condition, i.e., 
\begin{eqnarray}
P_{i_\alpha}(\infty)W_{i_\alpha j_\beta} \neq P_{j_\beta}(\infty)W_{j_\beta i_\alpha} . 
\end{eqnarray}	
The implies that the model under study is a non-equilibrium one that brings non-zero net probability currents in the steady state, defined as 
\begin{eqnarray}
f_{i_\alpha \to j_\beta}= P_{i_\alpha}(\infty)W_{i_\alpha j_\beta} - P_{j_\beta}(\infty)W_{j_\beta i_\alpha} .
\end{eqnarray}
It can be easily understood from the example shown in Fig.\ref{fig1}. On the one hand,  there must be some transitions that are microscopically irreversible, unless the topologies of each layer are the same. For example, the transition probability from node 1 in the first layer to node 3 in the second layer via one time-step (i.e., $1_1 \to 3_2$) is $\frac{1}{4} \times \pi_{12}$, while the reverse transition $3_2 \to 1_1 $ is impossible to occur in one time-step. The absence of microscopic reversibility breaks the detailed balance, as random walks on directed networks \cite{masuda2017random}. On the other hand, the net probability currents do not vanish even for reversible transitions (if $W_{i_\alpha j_\beta}>0$ then $W_{j_\beta i_\alpha}>0$, and vice versa).  We have calculated the net probability currents along all 17 reversible transitions in Fig.\ref{fig1}. Seven of them are shown in Fig.\ref{probflow}, from which two reversible transitions take place on the first layer, and two reversible transitions on the second layer, and the remaining three reversible transitions between the first layer and the second layer. For the sake of simplicity, we have set the transition probabilities between two layers to be equal, $\pi_{12}=\pi_{21}=p$. From Fig.\ref{probflow}, one sees that all net probability currents do not vanish for each $0<p <1$, and  thus verify the nonequilibrium nature of our model. } 

It was known that on a monoplex network the stationary occupation probability at each node equals to the normalized left eigenvector $\bm u^{(\alpha)}$ corresponding to the unit eigenvalue \cite{PhysRevLett.92.118701,masuda2017random}. For the standard random walk on networks, one has $\bm u^{(\alpha)}=d_i^{(\alpha)}/2 E^{(\alpha)}$, where $E^{(\alpha)}$ is the number of edges in the $\alpha$th layer. That is to say, the stationary occupation probability at each node on a monoplex network is proportional to the degree of the node. On the other hand, the stationary occupation probability at each layer is given by the left eigenvector of the transition matrix $\bm{\pi}$ between layers corresponding to the unit eigenvalue, i.e., $\bm{\mu} \pi= \bm{\mu}$ with $\bm{\mu}=(\mu_1,\cdots,\mu_\mathcal{L})$ satisfying $\sum_{\alpha=1}^{\mathcal{L}} \mu_\alpha=1$.

\begin{figure}
	\centerline{\includegraphics*[width=1.0\columnwidth]{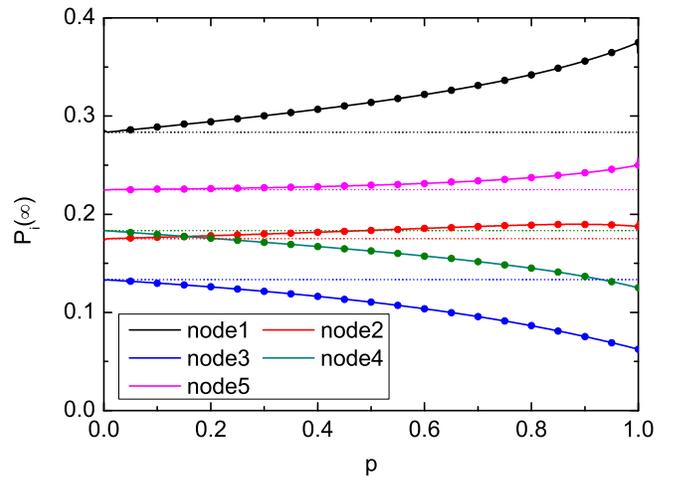}}
	\caption{The stationary occupation probability at each node on a multiplex network shown in Fig.\ref{fig1}. Here the transition probabilities between two layers are set to the same, $\pi_{12}=\pi_{21}=p$. Solid lines correspond to the theoretical values obtained from Eq.(\ref{eq5}) and Eq.(\ref{eq5.1}), dotted lines to the results from the perturbed theory in Eq.(\ref{eq22}), and solid circles to the results from Monte Carlo simulations.  \label{fig2}}
\end{figure}

An intuition is that the stationary occupation probability at each node on a multiplex network may be given by averaging the stationary occupation probabilities on monoplex networks, with the weight equals to the stationary probability $\mu_\alpha$ on each layer, i.e.,
\begin{eqnarray}\label{eq6}
{G_i}( \infty  ) = \sum\limits_{\alpha  = 1}^\mathcal{L} {\mu_\alpha }  \frac{d_i^{(\alpha)}}{2 E^{(\alpha)}},
\end{eqnarray} 
where we have used the letter ``G"  to avoid confusion with the results on multiplex networks. In fact, the intuition is untenable unless the transition probabilities between layers vanish. In Fig.\ref{fig2}, we show the stationary occupation probabilities at five nodes in Fig.\ref{fig1}. For the sake of simplicity, we have considered that the transition probabilities between two layers are equal, $\pi_{12}=\pi_{21}=p$, such that the walker will spend half of time on each layer on average. The results indicated by solid lines are obtained from power method to numerically solve the left eigenvector of $\bm{W}$ corresponding to the unit eigenvalue. The dotted lines correspond to the results in Eq.\ref{eq6}. We have also performed the Monte Carlo simulations, and the simulation results (see solid circles in Fig.\ref{fig2}) agree well with the theoretical ones in Eq.\ref{eq5}.  In the following, we will show by the perturbation theory that when the transition probabilities between layers are very small, $P_i(\infty)$ approximately equals to ${G_i}(\infty)$.

Without loss of generality, we only consider the case when the underlying network has two layers. The transition matrix between layers is given by
\begin{eqnarray}\label{eq7}
\bm{\pi}  = \left( {\begin{array}{*{20}{c}}
	{{\pi _{11}}}&{{\pi _{12}}} \\ 
	{{\pi _{21}}}&{{\pi _{22}}} 
	\end{array}} \right) = \left( {\begin{array}{*{20}{c}}
	{1 - {\pi _{12}}}&{{\pi _{12}}} \\ 
	{{\pi _{21}}}&{1 - {\pi _{21}}} 
	\end{array}} \right).
\end{eqnarray} 
The left eigenvector of $\bm{\pi}$  corresponding to the unit eigenvalue is simply given by $\bm{\mu}=(\frac{\pi_{21}}{\pi_{21}+\pi_{21}},\frac{\pi_{12}}{\pi_{12}+\pi_{21}})$.

\subsection{The limit $\pi_{12} \to 0$ and $\pi_{21} \to 0$}
For $\pi_{12}=\pi_{21}= 0$, the system can be decoupled into two subsystems, where the supra-Markov matrix can be written as $\bm{W}=\bm{W_0}$, where 
\begin{eqnarray}\label{eq11}
\bm{W_0} = \left( {\begin{array}{*{20}{c}}
	{\bm{W^{\left( 1 \right)}}}&{} \\ 
	{}&{\bm{W^{\left( 2 \right)}}} 
	\end{array}} \right)
\end{eqnarray}
is a block diagonal matrix. Let us denote by $\lambda^{(1)}_i$ ($\lambda^{(2)}_i$) the $i$th eigenvalue of $\bm W^{(1)}$ ($\bm W^{(2)}$), and by $\bm u _i^{( 1 )}$ ($\bm u _i^{( 2 )}$) and $ {\bm  v} _i^{( 1 )}$ ($ {\bm  v} _i^{(2) }$)  the corresponding left and right eigenvector, respectively. Since $\bm W^{(1)}$ and $\bm W^{(2)}$ are both Markov matrices, we have $1=\lambda_1^{(1)}>\lambda_2^{(1)}\geq\cdots\geq\lambda_N^{(1)} $ and $1=\lambda_1^{(2)}>\lambda_2^{(2)}\geq\cdots\geq\lambda_N^{(2)} $. The left and right eigenvectors corresponding to $\lambda_1^{(1,2)}=1$ are
$\bm{u}_1^{(1,2)}=\frac{1}{2 E^{(1,2)}} \left(d_1^{(1,2)}, \cdots, d_N^{(1,2)}\right)$ and $\bm{v}_1^{(1,2)}=\left(1, \cdots, 1\right)$, where $E^{(1,2)}$ is the number of edges in the first (second) layer. The eigenvalues of $\bm{W_0}$ are the set formed by the union of the eigenvalues of $\bm{W^{(1)}}$ and $\bm{W^{(2)}}$. They are $\left( {\lambda _1^{\left( 1 \right)}, \ldots ,\lambda _N^{\left( 1 \right)},\lambda _1^{\left( 2 \right)}, \ldots ,\lambda _N^{\left( 2 \right)}} \right)$, and the corresponding left eigenvectors are $\left( {\bm u_1^{( 1 )}, \bm 0} \right)$, $\left( {\bm u_N^{( 1 )}, \bm 0} \right)$, $\cdots$, $\left( {\bm 0, \bm u_1^{( 2 )} } \right)$, $\cdots$, $\left( {\bm 0, \bm u_N^{( 2 )} } \right)$, and the corresponding right eigenvectors are $\left( {\bm v_1^{( 1 )}, \bm 0} \right)$, $\left( {\bm v_N^{( 1 )}, \bm 0} \right)$, $\cdots$, $\left( {\bm 0, \bm v_1^{( 2 )} } \right)$, $\cdots$, $\left( {\bm 0, \bm v_N^{( 2 )} } \right)$, respectively.

In the limits, $\pi_{12} \to 0$ and $\pi_{21} \to 0$, we will use the perturbation theory to approximately calculate the left eigenvector of $\bm{W}$ corresponding to the unit eigenvalue, that is the stationary occupation probability of the walker on multiplex network. To the end, we first rewritten Eq.(\ref{eq4}) as
\begin{eqnarray}\label{eq10}
\bm{W} = \bm{W_0} + \Delta \bm{W},
\end{eqnarray}
where 
\begin{eqnarray}\label{eq12}
\bm{\Delta W} = \left( {\begin{array}{*{20}{c}}
	{ - {\pi _{12}}\bm{W^{\left( 1 \right)}}}&{{\pi _{12}}\bm{W^{\left( 1 \right)}}} \\ 
	{{\pi _{21}}\bm{W^{\left( 1 \right)}}}&{ - {\pi _{21}}\bm{W^{\left( 2 \right)}}} 
	\end{array}} \right)
\end{eqnarray}
is considered as a perturbed matrix. Since $\lambda=1$ is an eigenvalue of $\bm{W_0}$ with algebraic multiplicity two, we will use the degenerate perturbation theory to estimate the eigenvector corresponding to $\lambda=1$, as done in quantum mechanics. To the end, we first write the left eigenvector of $\bm{W}$ as a linear combination of the unperturbed system $\bm{W_0}$, i.e., 
\begin{eqnarray}\label{eq13}
\bm{u_1} = {c_1}\left( {\begin{array}{*{20}{c}}
	\bm{u_1^{( 1 )}} , 
	\bm{0} 
	\end{array}} \right) + {c_2}\left( {\begin{array}{*{20}{c}}
	\bm{0} , 
	\bm{u_1^{( 2 )}} 
	\end{array}} \right).
\end{eqnarray}
In terms of characteristic equation, 
\begin{eqnarray}\label{eq14}
 \bm{u_1} \left( \bm{W_0}  + \Delta \bm{W} \right)  = \left( {\lambda  + \Delta \lambda } \right) \bm{u_1}
\end{eqnarray}
Since $\bm{u_1}  \bm{W_0}    =  {\lambda  }  \bm{u_1}$, Eq.(\ref{eq14}) simplifies to
\begin{eqnarray}\label{eq15}
\bm{u_1} \Delta \bm{W}   = \bm{u_1} { \Delta \lambda } 
\end{eqnarray}
Right multiplying respectively by $\left( {\begin{array}{*{20}{c}}
	\bm{v_1^{( 1 )}} \\ 0 	\end{array}} \right)$ and $\left( {\begin{array}{*{20}{c}}
	0 \\ \bm{v_1^{( 1 )}} 	\end{array}} \right)$ both sides of Eq.(\ref{eq15}), we obtain homogeneous equations, 
\begin{eqnarray}\label{eq16}
\left( {\begin{array}{*{20}{c}}
	{{H_{11}} - \Delta \lambda }&{{H_{12}}} \\ 
	{{H_{21}}}&{{H_{22}} - \Delta \lambda } 
	\end{array}} \right)\left( {\begin{array}{*{20}{c}}
	{{c_1}} \\ 
	{{c_2}} 
	\end{array}} \right) = \left( {\begin{array}{*{20}{c}}
	0 \\ 
	0 
	\end{array}} \right),
\end{eqnarray}
where 
\begin{eqnarray}\label{eq17}
\begin{gathered}
{H_{11}} = {\left( {\begin{array}{*{20}{c}}
		{\bm u_1^{\left( 1 \right)}} , 
		\bm{0} 
		\end{array}} \right)}\Delta \bm {W}\left( {\begin{array}{*{20}{c}}
	{\bm v_1^{\left( 1 \right)}} \\ 
	\bm{0} 
	\end{array}} \right) =  - {\pi _{12}}, \hfill \\
{H_{12}} = {\left( {\begin{array}{*{20}{c}}
		\bm{0} , 
		{\bm u_1^{\left( 2 \right)}} 
		\end{array}} \right)}\Delta \bm {W}\left( {\begin{array}{*{20}{c}}
	{\bm v_1^{\left( 1 \right)}} \\ 
	\bm{0} 
	\end{array}} \right) = {\pi _{21}}, \hfill \\
{H_{21}} = {\left( {\begin{array}{*{20}{c}}
		{\bm u_1^{\left( 1 \right)}},
		\bm{0} 
		\end{array}} \right)}\Delta \bm {W}\left( {\begin{array}{*{20}{c}}
	\bm{0} \\ 
	{\bm v_1^{\left( 2 \right)}} 
	\end{array}} \right) = {\pi _{12}}, \hfill \\
{H_{22}} = {\left( {\begin{array}{*{20}{c}}
		\bm{0} ,
		{\bm u_1^{\left( 2 \right)}} 
		\end{array}} \right)}\Delta \bm {W}\left( {\begin{array}{*{20}{c}}
	\bm{0} \\ 
	{\bm v_1^{\left( 2 \right)}} 
	\end{array}} \right) =  - {\pi _{21}}. \hfill \\ 
\end{gathered} 
\end{eqnarray}
Nontrivial solutions for Eq.(\ref{eq16}) requires that
\begin{eqnarray}\label{eq18}
\det \left( {\begin{array}{*{20}{c}}
	{{H_{11}} - \Delta \lambda }&{{H_{12}}} \\ 
	{{H_{21}}}&{{H_{22}} - \Delta \lambda } 
	\end{array}} \right) = 0.
\end{eqnarray}
Substituting Eq.(\ref{eq17}) into  Eq.(\ref{eq18}), one obtain
\begin{eqnarray}\label{eq19}
{\Delta \lambda }_1=0 \quad \rm{or} \quad {\Delta \lambda }_2=-\pi_{12}-\pi_{21},
\end{eqnarray}
which leads to $\lambda_1=\lambda+{\Delta \lambda }_1=1$ and $\lambda_2=\lambda+{\Delta \lambda }_2=1-\pi_{12}-\pi_{21}$. We can seek that the perturbed left eigenvector corresponding to $\lambda_1=1$, which is the approximate stationary occupation probability. Substituting ${\Delta \lambda }_1=0$ into  Eq.(\ref{eq16}), we obtain
\begin{eqnarray}\label{eq20}
\left( {\begin{array}{*{20}{c}}
	{{c_1}} \\ 
	{{c_2}} 
	\end{array}} \right) \propto \left( {\begin{array}{*{20}{c}}
	{{\pi _{21}}} \\ 
	{{\pi _{12}}} 
	\end{array}} \right).
\end{eqnarray}
Substituting Eq.(\ref{eq20}) into  Eq.(\ref{eq12}) and then using the normalization condition, we obtain 
\begin{eqnarray}\label{eq21}
\bm{u_1} = \frac{1}{{{\pi _{12}} + {\pi _{21}}}}\left( {\begin{array}{*{20}{c}}
	{{\pi _{21}} \bm u_1^{( 1 )}} , 
	{{\pi _{12}}\bm u_1^{( 2 )}} 
	\end{array}} \right),
\end{eqnarray}
which gives the stationary occupation probability in the first-order perturbation theory, 
\begin{eqnarray}\label{eq22}
{P_i}\left( \infty  \right) = \frac{1}{{{\pi _{12}} + {\pi _{21}}}}\left( {\frac{{{\pi _{21}}d_i^{\left( 1 \right)}}}{{2{E^{\left( 1 \right)}}}} + \frac{{{\pi _{12}}d_i^{\left( 2 \right)}}}{{2{E^{\left( 2 \right)}}}}} \right).
\end{eqnarray}
We see that the stationary occupation probability in Eq.(\ref{eq22}) by the perturbation theory is the same as the result in Eq.(\ref{eq6}).

\subsection{The limit $\pi_{12}=\pi_{21}=1$}

In the opposite limit $\pi_{12}=\pi_{21}=1$, the supra-Markov matrix defined in Eq.(\ref{eq4}) is given by 
\begin{eqnarray}\label{eq8}
\bm{ W} = \left( {\begin{array}{*{20}{c}}
	\bm{0}&{ \bm {W^{\left( 1 \right)}}} \\ 
	{\bm {W^{\left( 2 \right)}}}&\bm{0} 
	\end{array}} \right).
\end{eqnarray}
Letting $\bm P = \left( {{{\bm P}^{( 1 )}}, {{\bm P}^{( 2 )}}} \right)$ be the stationary occupation probability in the limit, and using $\bm{P} \bm{{W}}=\bm{{P}}$, we have
\begin{eqnarray}\label{eq9}
\begin{gathered}
{{\bm P}^{\left( 1 \right)}} = {{\bm P}^{( 1 )}} \bm{W^{( 1 )}}\bm{W^{( 2 )}}, \hfill \\
{{\bm P}^{\left( 2 \right)}} = {{\bm P}^{( 2 )}}\bm{W^{( 2 )}}\bm {W^{( 1 )}}. \hfill \\ 
\end{gathered} 
\end{eqnarray}
This implies that ${\bm P}^{( 1 )}$ and  ${\bm P}^{( 2 )}$ are the left eigenvectors of the Markov matrices $\bm{W^{( 1 )}}\bm{W^{( 2 )}}$ and $\bm{W^{( 2 )}}\bm{W^{( 1 )}}$ corresponding to the unit eigenvalue, respectively. Generally speaking, ${\bm P}^{( 1 )}$ and  ${\bm P}^{( 2 )}$ are linearly independent, unless $\bm{W^{( 1 )}}$ and $\bm{W^{( 2 )}}$ are exchangeable in matrix multiplications.


\begin{figure*}
	\centerline{\includegraphics*[width=1.9\columnwidth]{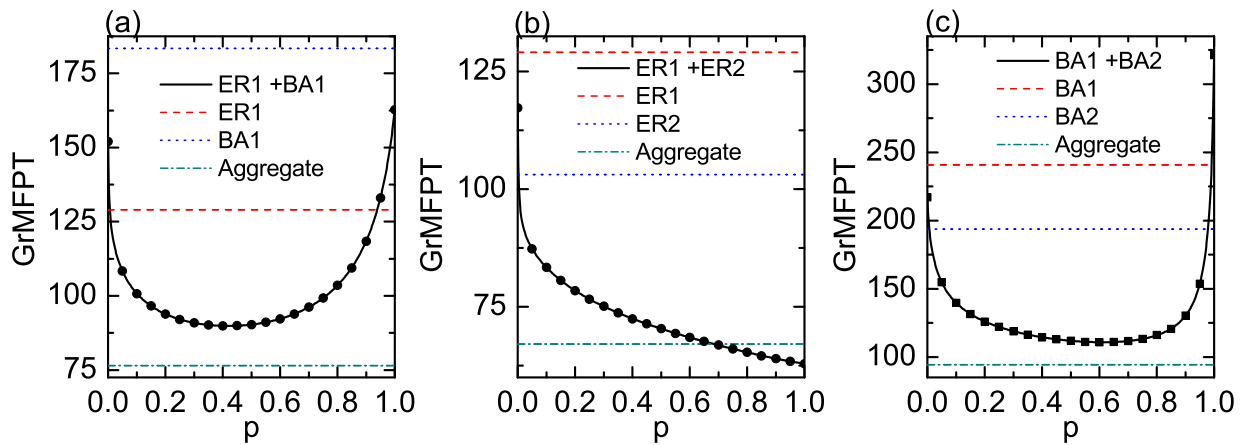}}
	\caption{The graph mean first passage time (GrMFPT) as a function of $p=\pi_{12}=\pi_{21}$ on three different multiplex networks, where each layer can be an Erd\"os–R\'enyi (ER) random network with the average degree $\langle k \rangle=3.48$ or a Barab\'asi–Albert (BA) scale-free network with the average degree $\langle k \rangle=2$. The size of network is fixed at $N=50$. Three horizontal lines indicate the GrMFPT on two monoplex networks and the aggregated network of two layers. Symbols correspond to simulation results from Monte Carlo. \label{fig4}}
\end{figure*}

\section{Mean First passage time}
Let us consider the first passage problem in a multiplex network. Assuming the walker starts from a node $i$ in the $\alpha$th layer, i.e., $i_{\alpha}$, how long does it arrive another node $j$ for the first time, regardless of the layer of the target node? Let us denote by $T_{i_\alpha,j}$ the MFPT from node $i_{\alpha}$ to node $j$ (any layer), which satisfies the following equation,
\begin{eqnarray}\label{eq30}
{T_{{i_\alpha }j}} = 1 + \sum\limits_{k \ne j} {{W_{{i_a}{k_\beta }}}} {T_{{k_\beta }j}} . 
\end{eqnarray}

Furthermore, the MFPT can be taken the average over the layer of starting node, with the weight proportional to the stationary occupation probability in each layer, given by
\begin{eqnarray}\label{eq31}
{T_{ij}} = \frac{{\sum\nolimits_{\alpha  = 1}^\mathcal{L} {{P_{{i_\alpha }}}\left( \infty  \right)} {T_{{i_\alpha }j}}}}{{\sum\nolimits_{\alpha  = 1}^\mathcal{L}  {{P_{{i_\alpha }}}\left( \infty  \right)} }}.
\end{eqnarray}

To measure the global search efficiency, we define the GrMFPT, which is the average of the MFPT  over all pairs of distinct nodes \cite{PhysRevE.89.012803},
\begin{eqnarray}\label{eq32}
{\rm{GrMFPT} }= \frac{1}{{N\left( {N - 1} \right)}}\sum\limits_{i \ne j} {{T_{ij}}} .
\end{eqnarray}

To illustrate our results, we consider different two-layer multiplex topologies,
where each layer can be an Erd\"os–R\'enyi (ER) random network \cite{ER1960} with the average degree $\langle k \rangle=3.48$ or a Barab\'asi–Albert (BA) scale-free network \cite{Science.286.509} with the average degree $\langle k \rangle=2$. For each layer, the size of network is fixed at $N=50$.   
For simplicity, we set $\pi_{12}=\pi_{21}=p$. In Fig.\ref{fig4}, we show the GrMFPT as a function of $p$ for three different multiplex networks. Interestingly, the GrMFPT shows a monotonic or nonmonotonic change with $p$. For the case of ER+BA or BA+BA topology, there exists an optimal value of $p$ for which the GrMFPT is a minimum. To validate the theoretical results, we have performed Monte Carlo simulations. The simulation results (see symbols in Fig.\ref{fig4}) obtained by averaging over $10^3$ realizations for each pair of distinct nodes, which agree well with theoretical ones.

For comparison, we also compute the GrMFPT on each monoplex network, given by two top horizontal lines in Fig.\ref{fig4}. The MFPT on each layer can be computed by numerically iterating the following equations,
\begin{eqnarray}\label{eq33}
T_{ij}^\alpha  = 1 + \sum\limits_{k \ne j} {W_{ik}^{(\alpha)} } T_{kj}^\alpha .
\end{eqnarray}	

Form Fig.\ref{fig4} , one finds that a wide range of $p$ exists for which the GrMFPT on the multiplex network is always less than those in two monoplex networks. This implies that the search on a multiplex network is more advantageous than any one of monoplex networks when the coupling between layers is properly set up.

We further consider the superposition of two-layer networks. The way to superposition is as follows. Two nodes $i$ and $j$ are considered to be connected only if they are connected in any layer, in the sense that the aggregate network is still unweighted and undirected. The entries of the adjacency matrix of such an aggregate network are given by
\begin{eqnarray}\label{eq34}
{A^S_{ij}} = \min \left\{ {1,\sum\limits_{\alpha=1}^{\mathcal{L}}  {A_{ij}^{(\alpha)} } } \right\}.
\end{eqnarray}	
The GrMFPT on the aggregate network is also shown in Fig.\ref{fig4} by the lowest horizontal line. We can see that for the case of ER+BA or BA+BA topology the GrMFPT on the multiplex network is larger than that in the aggregate network, but the minimal GrMFPT on the multiplex network is close to the GrMFPT in the aggregate network. While for the case of ER+ER topology, the GrMFPT on the multiplex is even smaller than that in the corresponding aggregate network when the transition probability between two layers is large enough.

\section{Comparison with other types of random walks}
\textcolor{blue}{In this section, we compare our model with two other types of random walks on multiplex networks, i.e., classical random walks (CRW) and maximal entropy random walks (MERW) \cite{PhysRevLett.102.160602}. For the CRW, the transition probability from node $i$ in the $\alpha$th layer to the node $j$ in the $\beta$th layer is defined as \cite{PNAS111.8351.2014} 
\begin{eqnarray}\label{eq4.1}
W_{{i_\alpha }{j_\beta }} = \delta_{\alpha \beta} \frac{A_{ij}^{(\alpha)}}{d_i^{(\alpha)}+D_i^{(\alpha)}}+(1-\delta_{\alpha \beta}) \frac{\delta_{ij} w_i^{\alpha \beta}}{d_i^{(\alpha)}+D_i^{(\alpha)}},
\end{eqnarray}
where $w_i^{\alpha \beta}=w_i^{\beta \alpha}$ denotes the connection weight between node $i$ in the $\alpha$th layer and its replica in the $\beta$th layer, and $D_i^{(\alpha)}=\sum_{\beta \neq \alpha} w_i^{\alpha \beta}$ is the strength of node $i$ with respect to connections
to its counterparts in different layers. The first term in Eq.(\ref{eq4.1}) indicates the probabilities of intralayer transitions, and the second term the probabilities of interlayer transitions.  }

\textcolor{blue}{Unlike CRW in which the transition of the walker depends on local information, the transition probability for the MERW is determined by the global structure of the network \cite{PhysRevLett.102.160602}. More specifically, the walkers
choose the next node to jump into maximizing the entropy
of their path at a global level, whereas classical random walkers maximize the entropy of their path at neighborhood level. The transition probabilities for the MERW is determined by the largest eigenvalue of the supra-adjacency matrix and the components of the corresponding eigenvector. 
The supra-adjacency matrix is defined as \cite{PNAS111.8351.2014}
\begin{eqnarray}\label{eq4.2}
\bm { A} = \left( {\begin{array}{*{20}{c}}
	{\bm{A^{\left( 1 \right)}}}& \cdots &{\bm{D^{\left( 1,\mathcal{L} \right)}}}\\
	\vdots & \ddots & \vdots \\
	{\bm{D^{\left( \mathcal{L}, 1 \right)}}}& \cdots &{\bm{A^{\left( \mathcal{L} \right)}}}
	\end{array}} \right),
\end{eqnarray}	
with ${\bm{D^{\left( \alpha, \beta \right)}}}={\rm diag} \left\{ w_1^{\alpha \beta}, \cdots, w_N^{\alpha \beta}  \right\}$.}

\textcolor{blue}{Finally, for the MERW the transition probability from node $i_\alpha$ to node $j_\beta$ is given by 
\begin{eqnarray}\label{eq4.3}
{W_{{i_\alpha }{j_\beta }}} &=& {\delta _{\alpha \beta }}\frac{{A_{ij}^{\left( \alpha  \right)}}}{{{\lambda _{\max }}}}\frac{{{\psi _{\left( {\alpha  - 1} \right)N + j}}}}{{{\psi _{\left( {\alpha  - 1} \right)N + i}}}} \nonumber \\ &+& \left( {1 - {\delta _{\alpha \beta }}} \right)\frac{{{\delta _{ij}}}}{{{\lambda _{\max }}}}\frac{{{\psi _{\left( {\alpha  - 1} \right)N + i}}}}{{{\psi _{\left( {\beta  - 1} \right)N + i}}}},
\end{eqnarray}
where $\lambda _{\max }$ is the largest eigenvalue of the supra-adjacency matrix $\bm{A}$, and $\psi _i$ is the $i$th component of the corresponding eigenvector. Once again, the first term in Eq.(\ref{eq4.3}) represents the probabilities of intralayer transitions, and the second term the probabilities of interlayer transitions.  }

\textcolor{blue}{For both CRW and MERW at the steady state, the detailed balance is satisfied, from which one can immediately obtain the stationary occupation probability, given by \cite{PNAS111.8351.2014}
\begin{eqnarray}\label{eq4.41}
{P_{{i_\alpha }}}\left( \infty  \right) = \frac{{d_i^{\left( \alpha  \right)} + D_i^{\left( \alpha  \right)}}}{{\sum_\beta  {\sum_j {\left( {d_j^{\left( \beta  \right)} + D_j^{\left( \beta  \right)}} \right)} } }}
\end{eqnarray}
for CRW, and 
\begin{eqnarray}\label{eq4.42}
{P_{{i_\alpha }}}\left( \infty  \right) = \psi _{\left( {\alpha  - 1} \right)N + i}^2.
\end{eqnarray}
for MERW.}

\begin{figure}
	\centerline{\includegraphics*[width=1.0\columnwidth]{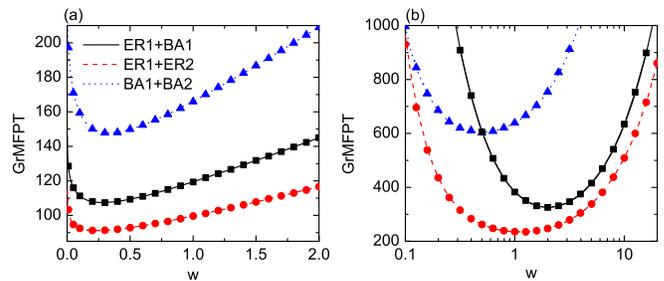}}
	\caption{\textcolor{blue}{The graph mean first passage time (GrMFPT) for CRW (a) and MERW (b) as a function of $w$ on three different multiplex networks, where each layer can be an ER random network with the average degree $\langle k \rangle=3.48$ or a BA scale-free network with the average degree $\langle k \rangle=2$. The size of network is fixed at $N=50$. $w$ denotes the strength of connections between different layers. Lines and symbols correspond to the results from solving Eq.(\ref{eq30}) and directedly from Monte Carlo simulations, respectively.} \label{fig5}}
\end{figure}

\textcolor{blue}{We now compute the GrMFPT defined in Eq.(\ref{eq32}) for CRW and MERW. To the end, we choose, for simplicity, $w_i^{\alpha \beta} \equiv w$ that is independent of the indices of node and layer. In Fig.\ref{fig5}(a) and Fig.\ref{fig5}(b), we show the GrMFPT as function of $w$ for CRW and MERW, respectively. We also used three different topologies of duplex, as considered in Fig.\ref{fig4}. Interestingly, for all cases the GrMFPT shows a unique minimum at some value of $w$. The minimum GrMFPT and the GrMFPT on the corresponding aggregated network are summarized in Table \ref{tb1}.}

\begin{table}
\centering
\caption{\textcolor{blue}{The minimal GrMFPT on multiplex networks and on the corresponding aggregated network for different types of random walks. Here ER and BA networks are the same as those in Fig.\ref{fig4} and Fig.\ref{fig5}.} }\label{tb1}
\begin{tabular}{ccccc}
 \toprule
	Multiplex & CRW & MERW & Our model & Aggregate \\ 
	 \midrule
	ER+BA & 107.5 & 325.7 & 89.9  & 76.5 \\ 
	ER+ER & 91.2 & 234.7 & 64.4 & 67.1 \\  
	BA+BA & 147.8 & 603.5 & 110.8  & 94.3 \\ 
	CS-Aarhus & 129.7 & 1005.8 & 161.2 & 95.6 \\
	London public transport & 3546.8 & 4265.5 & 3494.0 &3247.7  \\
	\bottomrule
\end{tabular}
\end{table}

\section{Application to real multiplex networks}
\begin{figure}
	\centerline{\includegraphics*[width=1.0\columnwidth]{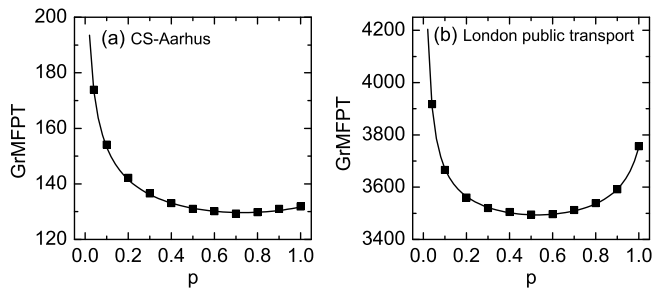}}
	\caption{\textcolor{blue}{The graph mean first passage time (GrMFPT) as a function of the transition probability $p$ between different layers on two real multiplex networks: (a) CS-Aarhus and (b) London public transport (see main text for details). Symbols correspond to simulation results from Monte Carlo.} \label{fig6}}
\end{figure}

\textcolor{blue}{Finally, we present the results on two real multiplex networks. The first example is a multiplex social network. It consists of five kinds of online and offline relationships (Facebook, Leisure, Work, Co-authorship, Lunch) between the employees of Computer Science department at Aarhus (CS-Aarhus for abbreviation). The multiplex network contains 61 nodes and 620 edges. The second example is the multiplex transportation
network of London (UK). There are 369 nodes and 441 edges in total. Nodes are train stations in London and edges encode existing routes between stations.
Underground, Overground and DLR stations are considered. We have computed the GrMFPT as a function of transition probability $p$ between different layers, as shown in Fig.\ref{fig6}. We can see that the GrMFPT shows a unique minimum at $p \approx 0.74$ and at $p \approx 0.52$ for the multiplex of CS-Aarhus and the multiplex of London public transport, respectively. The minimal GrMFPTs are summarized in Table \ref{tb1}. For comparison, we also computed the GrMFPT of CRW and MERW as a function of coupling strength between layers. The GrMFPT possesses a unique minimum at an intermediate level of coupling as well, and the minimum is presented in Table \ref{tb1}. For the multiplex of CS-Aarhus, the minimal GrMFPT of CRW is the smallest one. While for the multiplex of London public transport, the minimal GrMFPT for our model is the smallest one. 
For both real networks, the minimal GrMFPT of MERW is always the largest one.     }

\section{Conclusions}
To conclude, we have proposed a \textcolor{blue}{non-equilibrium} random walk model on multiplex networks. At each time step, the walker hops from one node to one of its neighbors in the same layer, as the standard random walk in a monoplex network, and then tries to switch between replicas of one node belong to different layers.
We have focused on the stationary occupation probability of the walker at each node and the GrMFPT. The former is given by the left eigenvector of the supra-Markov matrix corresponding to the unit eigenvalue. We show that the model at steady state does not satisfy the detailed balance and the stationary occupation probability does not equal to the superposition of counterparts in all layers. On the other hand, the GrMFPT shows a nontrivial dependence on the transition probability between different layers. In a wide range of parameter, the GrMFPT is smaller than that of any layer in isolation, and can even smaller than that of an aggregate network by all layers. 
\textcolor{blue}{Furthermore, we compare our model with two other types of random walks, i.e., classical random walks and maximal entropy random walks. Finally, we apply our model to two real multiplex networks. We found that for most of cases the minimum GrMFPT for our model is the smallest one among three models. Our results implies that the efficiency of random search on synthetic and real-world multiplex systems may benefit from the non-equilibrium nature of the dynamics. In the future, it is interesting to study the effect of topological overlap between layers on thermodynamic and/or dynamical quantities of ensembles of random walk trajectories, such as entropy production rate \cite{seifert2012stochastic} and large deviation of dynamical observables \cite{PhysRevE.99.022137}, which may be used to measure how far away from equilibrium on multiplex systems.   }      

\begin{acknowledgments}
This work is supported by the National Natural Science
Foundation of China (Grants No. 11875069, No. 61973001)
and the Key Scientific Research Fund of Anhui Provincial
Education Department (Grant No. KJ2019A0781).
\end{acknowledgments}


\end{document}